\documentclass[twocolumn,aps]{revtex4}%
\usepackage{graphicx}%
\usepackage{amsmath}%
\usepackage{amsfonts}%
\usepackage{amssymb}

\providecommand{\U}[1]{\protect\rule{.1in}{.1in}}
\begin{document}
\title{Dark polariton-solitons in semiconductor microcavities}
\author{A.V. Yulin$^{1,2}$, O.A. Egorov$^{1,3}$, F. Lederer$^{3}$, and D.V.
Skryabin$^{1}$}
\date{\today}
\address{
$^1$Centre for Photonics and Photonic Materials, Department of
Physics, University of Bath, Bath BA2 7AY, United Kingdom\\
$^2$Department of Engineering Mathematics, University of Bristol,
Bristol, BS8 1TR, United Kingdom\\
$^3$ Institute of Condensed Matter Theory and Solid State Optics,
Friedrich Schiller University Jena, Max-Wien-Platz 1, 07743 Jena,
Germany}

\begin{abstract}
We report the existence, symmetry breaking and other instabilities
of dark  polariton-solitons in semiconductor microcavities operating
in the strong coupling regime. These  half-light half-matter
solitons are potential candidates for applications in all-optical
signal processing. Their excitation time and required pump powers
are a few orders of magnitude less than those of their weakly
coupled light-only counterparts.
\end{abstract}

\maketitle

\input{epsf.tex} \epsfverbosetrue

\narrowtext Polaritons are mixed states of photons and material
excitations and are well-known to exist in many condensed matter,
atomic and optical systems \cite{book,book1,book2,nature0,Weis92}.
We are dealing below with a semiconductor microcavity, where
polaritons exist due to mixing of quantum well
excitons and resonant microcavity photons
\cite{book2,nature0,Weis92}. In the strong coupling regime photons,
emitted as a result of electron transitions, excite the medium and
are re-emitted in a cascaded manner, which gives rise to so-called
Rabi oscillations \cite{book,book2,nature0}. This phenomenon results
in the two peak structure of the microcavity absorption spectrum.
The measured spectral width of the peaks corresponds to the
picosecond polariton life time \cite{book2}. This is in contrast
with the more usual weak-coupling regime (typical for operation of
vertical cavity surface emitting lasers (VCSELs) \cite{book2}),
where the slow (nanosecond) carrier dynamics does not catch up with
the fast (picosecond) photon decay. Thereby most of the photons
leave the cavity as soon as they are emitted. In this regime the
response to a pulse, resonating with a cavity mode, results in a
single spectral peak. Thus any potential application of microcavity
polaritons  in optical information processing leads to a 2-3 orders
of magnitude response time reduction relative to the VCSEL-like
operating regimes.

One of the topics of the recent research into
the weakly coupled semiconductor microcavities has been the localised
structures of light or cavity solitons
\cite{old2,Tara01b,nature_sol,firth_apl,Tang08,Lari05,Pesh03}, which
have demonstrated rich physics and have  been
proposed for information processing applications
\cite{nature_sol,Tang08}. In
the weak coupling regime formation of polaritons is irrelevant,
since the dispersion of linear excitations is purely photonic.
Slowness of the light-only cavity solitons is an outstanding problem,
which can be rectified in the strong-coupling regime, where
potentially much faster, but not yet reported, light-matter solitons are expected.

In the last few years extensive studies of the polaritons in
strongly coupled microcavities  have  been strongly motivated by the
smallness of the polariton mass leading to observation of the polariton Bose-Einstein
condensation at few Kelvin temperatures \cite{Kasp06,trap}. Polaritons
have also been recently observed even at the room temperatures, see, e.g., \cite{room},
which has further boosted their potential for practical applications.
Another very important feature of
polaritons in semiconductor microcavities is their strong repulsive
interaction (two-body scattering) resulting in a substantial
defocusing nonlinearity \cite{book1,book2}. Amongst  nonlinear
effects predicted or observed with microcavity polaritons are
optical bistability \cite{Tred96,Baas04,Gippius04,krizh,recent} and parametric conversion
\cite{Steven2000,Ciut03,Gippius04,Caru04,nature11}. Observation of
these effects with polaritons requires pump intensities of $\sim
100\mathrm{W/cm}^{2}$ or below (see, e.g., Fig. 1 in \cite{Baas04}),
which is less than the typical pump  of $10\mathrm{kW/cm}^{2}$
required for semiconductor microcavities operating in the
weak-coupling regime
\cite{old2,Tara01b,nature_sol,firth_apl,Tang08,Lari05} (see, e.g.,
Fig. 5 in \cite{Lari05}).

Solitonic effects with polaritons in bulk media have attracted a
significant (mostly theoretical) attention since  70s till now,
see, e.g., Ref. \cite{book1,maim}.   In
the latest wave of research on exciton-polaritons in strongly
coupled microcavities the solitonic effects have not been much of a
focus yet, with an important exception of a recent experimental
paper \cite{Lari08}. In this work the authors claim observation of
dark and bright localized structures or cavity solitons in a
strongly coupled semiconductor microcavity. Some other papers have
reported localisation of microcavity polaritons due to linear
defects \cite{trap,kav2}, as a result of switching between two
polarizations \cite{kav1}, or neglecting such important requisites
of passive cavities as losses, external pump and hence bistability
\cite{kam}. For studies of spatially dependent polariton dynamics,
see, e.g., \cite{kav10}. Our work is aimed at filling an existing
gap in the theoretical knowledge about microcavity polariton-solitons. This
is necessary not only for backing so far limited experimental
observations \cite{Lari08}, but also and mainly for guiding the
future work in this direction.

The widely accepted dimensionless mean-field model for excitons
strongly coupled to the circularly polarized cavity photons is
\cite{book1,book2,Caru04}
\begin{align}
& \partial_{t}E-i(\partial_{x}^{2}+\partial_{y}^{2})E+(\gamma_{c}%
-i\Delta)E=E_{p}+i\Psi,\nonumber\\
& \partial_{t}\Psi+(\gamma_{0}-i\Delta+i|\Psi|^{2})\Psi=iE.\label{e1}%
\end{align}
Here $E$ and $\Psi$ are the averages of the
photon and exciton creation or annihilation  operators.
Normalization is such that $(\Omega_R/g)|E|^2$ and
$(\Omega_R/g)|\Psi|^2$ are the photon and exciton numbers per unit
area. Here,  $\Omega_R$ is the Rabi frequency and $g$ is the
exciton-exciton interaction constant.
$\Delta=(\omega-\omega_r)/\Omega_R$ describes detuning of the pump
frequency $\omega$ from the identical resonance frequencies of
excitons and cavity, $\omega_r$. Time $t$ is measured in units of $1/\Omega_R$.
$\gamma_c$ and $\gamma_0$ are the
cavity and exciton damping constants normalized to $\Omega_R$.
Transverse coordinates $x$, $y$ are normalized to the value
$x_0=\sqrt{c/2kn\Omega_R}$ where $c$ is the vacuum light velocity,
$n$ is the refractive index and $k=n\omega/c$ is the wavenumber.
The normalized amplitude of the external pump $E_p$
is related to the physical incident  intensity $I_{inc}$ as
 $|E_p|^2=g\gamma_cI_{inc}/\hbar\omega_0{\Omega_R}^2$
\cite{Caru07}. As a guideline for realistic estimates one can use parameters
 for a  microcavity with a single InGaAs/GaAs
quantum well: $\hbar\Omega_R\simeq 2.5meV$,
$\hbar g\simeq 10^{-4}eV\mu m^2$, see \cite{Baas04,Caru04,Caru07}. Assuming the
relaxation times of the photonic and excitonic fields to be
 $2.5ps$ gives $\gamma_{c,0}\simeq 0.1$. In accordance with this set of parameters the
normalized driving amplitude $|E_p|^2=1$ physically corresponds to
the external pump intensity  $\sim 10kW/cm^2$.
Optical bistability appears for $|E_p|\sim 0.1$, it gives the input intensity
$\sim 100 W/cm^2$. Experimentally the polariton bistability has been observed for values close or even less than
$100 W/cm^2$ \cite{Tred96,Baas04,Gippius04,krizh,recent}.

\begin{figure}[ptb]
\setlength{\epsfxsize}{2.0in}
\centerline{\epsfbox{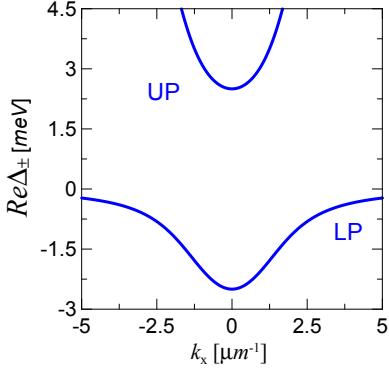}}\caption{Polariton dispersion
calculated from Eq. (\ref{disp}), $Re\Delta_{\pm}(k_{x})$, and
renormalized back into physical units. Parameters are $n=3.5$,
operating wavelength $\lambda=0.85\mu$m,
$\gamma_{c,0}=0.1$, $\hbar\Omega_{R}=2.5$meV.}%
\label{disp_ch}%
\end{figure}

First we briefly summarize important aspects of the linear dispersion and
bistability properties of the above equations. Assuming that $E,\Psi\sim
e^{ik_xx+ik_yy}$ and neglecting pump and nonlinearity we find the
dispersion law of cavity polaritons
\begin{equation}
\Delta_{\pm}={\frac{k^{2}-i(\gamma_{c}+\gamma_{0})}{2}}\pm\sqrt{1+{\frac{(k^{2}+i(\gamma_{0}-\gamma_{c}))^{2}}{4}}},\label{disp}%
\end{equation}
where $k^2=k_x^2+k_y^2$. $Re\Delta_{+}$ corresponds to the frequency of the upper polariton
(U-polariton) and $Re\Delta_{-}$ to the lower polariton (L-polariton)
branch, see Fig.~\ref{disp_ch}. In the strong coupling regime the gap between
U- and L-polaritons is greater than the linewidth of the branch due
to $Im\Delta_{\pm}\neq0$.

\begin{figure}[ptb]
\setlength{\epsfxsize}{3.in}
\centerline{\epsfbox{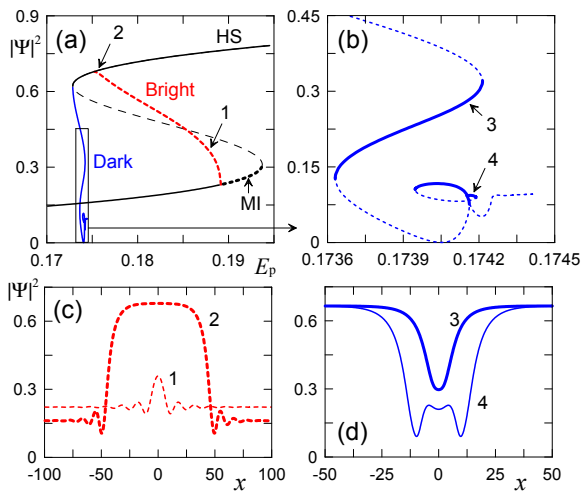}}\caption{ (a) Amplitude of the
homogeneous state (HS) (black line), $\max|\Psi(x,y)|^2$
for bright solitons  (blue line) and
$\min|\Psi(x,y)|^2$ for dark solitons (red line) shown as functions of
$E_p$: $\Delta=-0.7$, $\gamma_{0,c}=0.1$. (b) is the zoom of the
rectangular area from (a) showing bifurcations of the dark solitons.
(c,d) Exciton density  distribution $|\Psi(x,y=0)|^{2}$ across the
bright (c) and dark (d) solitons  for the points marked by 1, 2, 3 and 4.
Full and dashed lines in (a)-(d) mark stable
and unstable solutions, respectively. }
\label{ing_low_st}%
\end{figure}

If $E_{p}\neq0$, then solitons can exist only on a finite
amplitude background ($E(\pm\infty)\ne 0$), simply because the zero homogeneous solution is absent.
Therefore we proceed with a brief consideration of spatially homogeneous
solutions (HSs) and their stability. Then we report the existence of various
 cavity polariton solitons (CPSs) and
study their stability and instability scenarios. HS having  bistable dependence
 from $E_{p}$ is an important prerequisite
for the soliton existence. $E(E_{p})$ is multivalued provided that $f(\Delta)>0$,
where
\begin{equation}
f(\Delta)\equiv\Delta(\Delta^{2}+\gamma_{c}^{2}-1)-\sqrt{3}%
\gamma_{0}(\Delta^{2}+\gamma_{c}^{2}+{\frac{\gamma_{c}}{\gamma_{0}}}).
\end{equation}
The cumbersome expressions for the roots of $f(\Delta)=0$ simplify
for $\gamma_{c}=\gamma_{0}=0$ and give two bistability intervals
$\Delta>1$ and $-1<\Delta<0$. These two intervals overlap with the
$\Delta$ intervals allowed by the dispersion relation, see Eq.
(\ref{disp}) and Fig. 1. The bistability in the interval
$-1<\Delta<0$ appears because of the nonlinear resonance of the pump
with the L-polaritons whereas the bistability in the semi-infinite
interval is associated with the nonlinear resonance of the pump with
the U-polaritons.  Weakly coupled
cavities with defocusing nonlinearities exhibit bistability only for
$\Delta>0$, see, e.g. \cite{Pesh03}. Below we focus our attention on
the solitons linked to L-polaritons,
therefore our studies are unique to the strong coupling regime.  Stability analysis of the HS
L-polaritons
($\Delta<0$) has been previously reported for example in Refs.
\cite{Caru04}. The lower state of the L-polariton bistability loop
can be modulationally unstable within some interval of
$E_{p}$, while the upper state is generally stable, see Figs.~\ref{ing_low_st}%
(a) and 3(a). Here, modulational instability (MI) we understand as
the growth of linear perturbations in the form $e^{ik_xx+ik_yy+\kappa t}$ ($Re\kappa$ is the
growth rate).
As $\Delta$ is changing from the bottom of the L-polariton branch
towards the linear exciton resonance, $\Delta=0$, the point of MI is moving towards the left edge
of the bistability loop and finally goes beyond the latter, cf. Figs. 2(a) and 3(a).

Restricting ourselves to the structures independent on the polar angle ($\theta=arg(x+iy)$) we find that the
time-independent CPSs obey
\begin{equation}
-i\left(  {\frac{d^{2}E}{dr^{2}}}+{\frac{1}{r}}{\frac{dE}{dr}}\right)
+(\gamma_{c}-i\Delta)E=E_{p}+i\Psi\label{e4}%
\end{equation}
where $r=\sqrt{x^2+y^2}$, $\Psi=iE/[(\gamma_{0}-i\Delta)+iz]$ and
$z\equiv|\Psi
|^{2}$ is found solving the real cubic equation $(\gamma_{0}^{2}%
+(z-\Delta)^{2})z=|E|^{2}$. $z$ turns out to be a single valued
function of $|E|^{2}$ throughout the range of parameters
corresponding to the bistability of L-polaritons. Thus the potential
problem of ambiguity in choosing a root for $z$ is avoided.

We start our analysis of cavity polariton solitons from the case,
when the MI point of low state L-polaritons is within the bistability
interval. In many previously studied models bifurcation points of the
homogeneous solutions  have been the sites where localized structures
 branch off \cite{Pesh03}. Applying the Newton iterative method to Eq.
(\ref{e4}) we have found a family of small amplitude bright CPSs emerging from
the MI point, see the dashed red line in Fig.~\ref{ing_low_st}(a). Going
towards smaller values of $E_{p}$, the CPSs become more intense, see
Fig.~\ref{ing_low_st}(c). The $E_{p}$ value, at which the lower and upper
homogeneous states can be connected by a standing 1D front, is called Maxwell
point and this is the point where the branch of the bright CPSs terminates
($E_{p}=0.1748$).
When the pump approaches the Maxwell point the soliton broadens and its peak intensity tends towards
the intensity of the upper homogeneous state.
We also perform a full 2D linear stability analysis of the found
structures. The linear perturbations are assumed in the general form
$\epsilon_{+}(r)e^{iJ\theta+\kappa t}+\epsilon_{-}^{\ast}(r)e^{-iJ\theta
+\kappa^{\ast}t}$, where  $J=0,1,2,\dots$ \cite{skr}.
The resulting Jacobian operator is analysed using finite differences in $r$.
The linear stability analysis shows that the bright CPSs are unstable with
respect to the perturbation with the azimuthal index $J=0$ and that the
development of the instability splits the CPS into 2D moving fronts. When
$E_{p}$ is close to the Maxwell point this instability is relatively weak and
bright CPSs can be easily stabilized by the spatial inhomogeneities of the pump or cavity detuning. This
problem deserves more detailed investigation and it will be analyzed elsewhere.

\begin{figure}[ptb]
\setlength{\epsfxsize}{3.in}
\centerline{\epsfbox{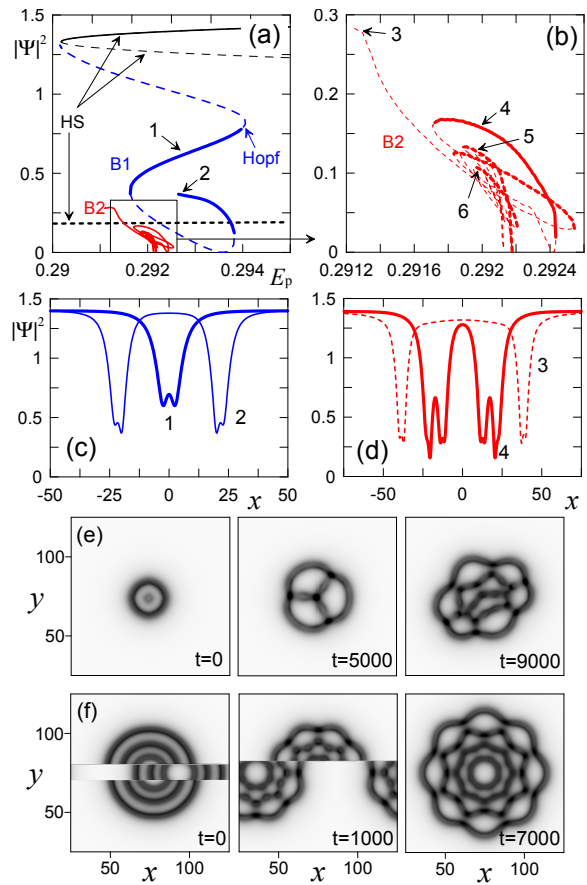}}\caption{ a) Amplitude  of the homogeneous state (HS)
 (black line) and $\min|\Psi(x,y)|^2$ for dark solitons (red
and blue lines) shown as functions of  $E_p$: $\Delta=-0.5$,
$\gamma_{0,c}=0.1$. $B1$ and $B2$ mark two branches of dark CPSs.
(b) is the zoom of the rectangular area from (a) showing
bifurcations of the B2 dark CPSs. (c,d) Exciton density distribution
$|\Psi(x,y=0)|^{2}$ across B1 (c) and B2 (d) CPSs  for the points
marked by 1, 2, 3 and 4 in panels (a) and (b). Full and dashed lines
in (a)-(d) mark stable and unstable solutions, respectively. (e,f)
show development of the symmetry breaking instabilities of the CPSs
marked as 5 and 6 in (b). }
\label{ing_low_ust}%
\end{figure}

Because of the defocusing nature of the polaritonic nonlinearity dark CPSs,
see, e.g., \cite{Pesh03}, are expected to be naturally selected by our system
and the instability of bright CPSs is not surprising. Dark cavity solitons, have
been previously studied both theoretically and experimentally for
semiconductor microcavities in the weak-coupling regime, see, e.g.
\cite{Mich98,Lari05,Egor07}. Unlike fiber solitons, the dark cavity
solitons have no conceptual disadvantage over the bright ones as information carriers.
The branch of dark CPSs have been found to detach from the left
folding point of the bistability loop and tend towards the Maxwell point, see
Fig.~\ref{ing_low_st}(a) and the zoomed area in (b).  At the onset of their existence
the dark solitons are seen only as a very deviation from the homogeneous background. As
$E_{p}$ tends towards the Maxwell point from the right, they become much
dipper. Near the Maxwell point the dark solitons become very broad
and can be roughly considered as superpositions of
infinitely separated  1D fronts ($1/r$ term in Eq. (\ref{e4}) can
be disregarded for large distances and the equation becomes effectively 1D).
It is important to note that the
relaxation of the fronts towards the upper state happens without oscillations,
however the relaxation towards the lower state is  oscillatory, see
Fig.~\ref{ing_low_st}(c). Thus pinning of the two fronts and hence
stabilization of CPSs is possible only for the dark structures (see the thick line in Fig. 3(c)).
The stable branches of dark CPSs are shown by full
lines in Fig.~\ref{ing_low_st}(b). The unstable ones correspond to the
instabilities with $J=0$.

In the case when the lower branch HS is unstable within the whole range of the
bistability ($\Delta=-0.5$) we have not found bright solitons, see Fig. 3.
This result is not surprising because the
 bright soliton branch is expected to detach from the point where
the lower HS changes its stability. This point is now well out of
the bistability range, which is another prerequisite for their existence. However, we have
found two distinct branches of dark CPSs marked as B1 and B2 in
Figs.~\ref{ing_low_ust}(a),(b). The B1 branch bifurcates subcritically from
the folding point of the upper homogeneous state. Initially unstable ($J=0$)
CPSs become stable after the turning point. Close to this turning
point the B1 CPSs have a deep\textbf{ }like shape, while later they transform
into dark rings of growing radius Fig.~\ref{ing_low_ust}(c).
Note, that close to the turning points additional destabilization of dark CPS
happens  due to linear eigenmodes with complex  $\kappa$ and $J=0$ (Hopf instability,
see Fig.~\ref{ing_low_ust}(a))
resulting in the formation of oscillating dark CPSs.

The branch B2 consists of ring shaped
structures, see Fig.~\ref{ing_low_ust}(d).
The linear
stability analysis shows that the B2 CPSs can be stable (see the interval
marked by $4$ in Fig.~\ref{ing_low_ust}(b)). However, more often, they are
unstable with respect to perturbations breaking the radial symmetry, i.e. with
$J\neq0$. An example of this instability development is shown in
Figs.~\ref{ing_low_ust}(e,f). The dark ring CPS shown in Fig. 3(e) is unstable
against linear eigenmode with $J=3$. The broader CPSs undergo azimuthal
instabilities with larger azimuthal numbers $J$. For example, $J=8$ for the
concentric ring CPS shown in Fig.~\ref{ing_low_ust}(f).

In summary: Following a series of recent experiments on observation of microcavity
polaritons, we have studied the formation of spatially localised
polariton-soliton structures in the strong coupling regime. In particular, our results can be
used for the interpretation of the experimental measurements reported in
\cite{Lari08}, where the polariton Rabi splitting has been observed
simultaneously with the formation of various bright and dark localised
structures. Microcavity polariton solitons
reported here exhibit a picosecond excitation time and can be
observed at pump powers few orders of magnitude lower than those
required in the weak coupling regime of the semiconductor microcavities
\cite{old2,Tara01b,nature_sol,firth_apl,Tang08,Lari05}. Thus the light-matter  polariton solitons
have potentially significant advantages in all-optical signal processing applications
over the light-only cavity solitons \cite{nature_sol,firth_apl,Pesh03}.


\end{document}